\documentstyle[preprint,aps]{revtex}
\begin{document}
\draft
\title{Correlated charge polarization  
in a  chain of coupled quantum dots}
\author{R. Kotlyar$^{1}$, C. A. Stafford$^{2}$, and S. Das Sarma$^{1}$}
\address{$^{1}$Department of Physics,
University of Maryland, College Park, MD 20742-4111, USA\\
$^{2}$Fakult\"at f\"ur Physik, Albert-Ludwigs-Universit\"at, D-79104 Freiburg, 
Germany} 
\date{\today}
\maketitle

\begin{abstract}
Coherent charge transfer in a linear array of tunnel-coupled quantum dots,
electrostatically coupled to external gates, is investigated using the Bethe
ansatz for a symmetrically biased Hubbard chain.  Charge polarization in
this correlated 
system is shown to proceed via two distinct processes:  formation of
bound states in the metallic phase, and charge transfer processes 
corresponding to a superposition of antibound states at opposite
ends of the chain in the Mott-insulating phase.
The polarizability in the insulating phase of the chain exhibits a 
universal scaling behavior, while the polarization charge
in the metallic phase of the 
model is shown to be quantized in units of $e/2$.
\end{abstract}
\pacs{PACS numbers: 73.20.Dx, 73.23.Hk, 71.10.Fd}

\widetext
\tighten

Tunneling of a single electron from one region to another in a mesoscopic
system leads to a modification of the dielectric response of the system
\cite{markusandi} which can be detected via single-electron capacitance
spectroscopy \cite{ashoori}.  
Capacitance measurements allow one to 
study charge transfer {\em in equilibrium}, and thus provide an important 
alternative to transport measurements 
\cite{dotarrays,waugh,chargetransfer,yacoby} in probing the
effects of coherent tunneling.  In this Letter, we investigate the charge 
polarization of a linear array of 
tunnel-coupled quantum dots embedded between the plates of a capacitor
(Fig.\ 1).   The quantum corrections to the classical two-terminal capacitance
of the system are shown to exhibit sharp resonances whose structure reveals
directly the spatial correlations of the interacting 
many-body ground state of the system.
We find that the localized character of the many-body states in the 
Mott-insulating phase of the model leads to extremely sharp capacitance
resonances, which obey a universal scaling form analogous to the conductivity
of the system \cite{andyandi}. On the other hand, the extended quantum
states in the metallic phase of the model are shown to lead to fractional
charge transfer, in contrast to the integer charge transfer predicted
in Ref.\ \cite{markusandi}. 

The integrability of 1D quantum many-body systems with open boundary conditions
was first established \cite{schulz} for the one-dimensional (1D)
Hubbard model.  Bethe ansatz solution was recently extended
to include boundary potentials \cite{suzuki}, and
the spectrum of bound states for a single attractive boundary potential
has very recently been investigated \cite{frahm}.  
Here, we investigate
a Hubbard chain with equal and opposite boundary potentials at each end,
which serves as a model of a capacitively biased
1D array of quantum dots.  In addition to the bound states found for
the case of a single boundary potential \cite{frahm}, we find
charge transfer states, which are quantum mechanical
superpositions of antibound states at opposite ends of the chain.
These charge transfer states are shown to dominate the 
polarizability in the Mott-insulating phase of the model.

We consider a closed linear system of quantum dots coupled electrostatically to
bias gates and a backgate (Fig.\ 1).  The backgate allows the system to be
charged with $N$ excess electrons,  
this excess charge being shared among the dots in 
the chain by quantum-mechanical tunneling.
We describe this coupled quantum dot chain  by the Hubbard 
model \cite{stafsds}
in the experimentally accessible limit when the interdot capacitances are 
negligible 
compared to the capacitances $C_g$ to the external gates.
The Hamiltonian of the system, 
including the work done by the external voltage sources, is
\begin{equation}
H  =  -t\sum_{\sigma}\sum_{i=1}^{L-1} \left(c_{i+1\sigma}^{\dagger}
 c_{i\sigma} + \mbox{H.c.}\right) 
+ \frac{U}{2} \sum_{i=1}^{L} \rho_i^2 
-C_0 V^2/2 + \frac{eV}{2}\left(\rho_L-\rho_1\right),
\label{hubham}
\end{equation}
where $c_{i\sigma}^{\dagger}$ creates an electron of spin $\sigma$ on
dot $i$, $\rho_i=\sum_{\sigma} c_{i\sigma}^{\dagger} c_{i\sigma}$, 
$U=e^2/C_g$ is the charging energy of a quantum dot,  
and $C_0$ is the classical
geometrical capacitance between the left and right gates.  Eq.\ (1)
is the prototypical minimal model of correlated fermions on a lattice, and
describes {\em e.g.}, the correlation-induced metal-insulator
transition \cite{andyandi}.  The new feature investigated
here is the nonperturbative effect of the external bias ($V$) described 
by the last term in Eq.\ (1), which polarizes the system.
Unlike previous investigations of the charge response of the 
system \cite{andyandi}, we
do not treat the bias $V$ as a weak perturbation, but consider arbitrarily
large values of $V$, leading to a finite transfer of charge across the chain.  
The polarization charge $Q$ induced on the external capacitor
plates characterizes the measurable dielectric response of the system.
At zero temperature,
the expectation value of the polarization charge 
is given by
\begin{equation}
\langle Q\rangle=\langle Q_{L}-Q_{R}\rangle/2=
-  \partial E_0/\partial V, 
\label{QandC.def}
\end{equation}
where $Q_{L}$ ($Q_{R}$) is the polarization charge on the 
left (right) capacitor plate and $E_0$ is the minimum eigenvalue of 
Eq.\ (1). The two-terminal capacitance  of the device 
is defined as  $C_{\mu} = -\partial^2 E_0/\partial V^2$. 
These quantities can be exactly obtained for the quantum dot chain using 
the Bethe ansatz technique, as described below.

The eigenvalues of Eq.\ (1) may be expressed as
\begin{equation}
E= -2t\sum_{j=1}^N \cos k_j -C_0V^2/2,
\label{bethe.ener}
\end{equation}
where the pseudomomenta $k_j$ are a set of $N$ distinct numbers which satisfy
the coupled equations
\begin{equation}
S_V(k_j)\, e^{i2k_j(L+1)}
= \prod_{\beta=1}^M 
\frac{\sin k_j -\lambda_{\beta}+iU/4t}{\sin k_j -\lambda_{\beta}-iU/4t}\,
\frac{\sin k_j +\lambda_{\beta}+iU/4t}{\sin k_j +\lambda_{\beta}-iU/4t},
\label{bethe.k}
\end{equation}
\begin{equation}
\prod_{j=1}^N 
\frac{\lambda_{\alpha}-\sin k_j +iU/4t}{\lambda_{\alpha}-\sin k_j -iU/4t}\,
\frac{\lambda_{\alpha}+\sin k_j +iU/4t}{\lambda_{\alpha}+\sin k_j -iU/4t}
=
\prod_{\beta(\neq \alpha)=1}^M 
\frac{\lambda_{\alpha}-\lambda_{\beta}
+iU/2t}{\lambda_{\alpha}-\lambda_{\beta}-iU/2t}\,
\frac{\lambda_{\alpha}+\lambda_{\beta}
+iU/2t}{\lambda_{\alpha}+\lambda_{\beta}-iU/2t},
\label{bethe.lambda}
\end{equation}
where $\lambda_{\alpha}$, $\alpha=1,\ldots,M$ are a set of distinct numbers
referred to as spin rapidities, and 
\begin{equation}
S_V(k_j)=\frac{1-(eV/2t)^2 e^{-2ik_j}}{1-(eV/2t)^2 e^{2ik_j}} 
\label{s.boundary}
\end{equation}
is the single-electron scattering matrix associated with the boundary
potentials.

The capacitive response of a chain of 4 quantum dots, calculated from
Eqs.\ (2--6), is shown in Fig.\ 1 for
several values of $N$.
The polarization induced by the external bias $V$ leads to a transfer of
charge across the system, which is reflected in the appearance of complex
roots of the Bethe ansatz equations, corresponding to bound and antibound
states on the boundary dots (see Table I).
Let us first consider the Mott-insulating phase of the system, which occurs
\cite{andyandi,stafsds} for commensurate electron density, $N=L$.
For low bias, $eV< 2t$, the Bethe ansatz ground state contains only real
pseudomomenta, and the charge distribution remains essentially symmetric.
For $2t<eV \lesssim U$, a bound state forms on the leftmost dot, characterized
by the complex pseudomomentum $k_L$.  However, due to the incompressibility
of the Mott insulator, an antibound state on the rightmost dot is also
filled ($k_{L-1}$), and there is thus no net transfer of charge.  
The Mott-Hubbard gap is reflected in the 
suppression of the low-bias capacitance (the dash-triple-dotted curve in Fig.\
1).  For a bias larger than the Mott-Hubbard gap, however,  it becomes 
energetically favorable to depopulate the antibound state on the rightmost
dot and populate an antibound state in the upper Hubbard band
on the leftmost dot (region II in Table I).  The pseudomomentum of this
antibound state   
contributes $-2t\cos k_{L-1} = \left[(U-eV/2)^{2}+4t^2\right]^{1/2}
\simeq U-eV/2$ to the ground state energy in Eq.\  (3)
(plus small backflow terms), indicating the presence of a second electron
on the leftmost dot.
The resulting
transfer of an electron across the array leads to a sharp capacitance 
resonance at $eV=eV_{1}\simeq U$ in Fig.\ 1.  Finally, for $eV > eV_{2} \simeq
2U$, this antibound state becomes a bound state.

In order to elucidate the nature of the charge-transfer resonance in the
Mott insulator, let us
first consider the simplest case $L=2$,
for which Eq.\ (1) reduces to a simple $4\times 4$ matrix.
The polarization charge and capacitance may then be obtained directly 
[neglecting terms of order $(t/U)^2$],
\begin{equation}
\frac{Q - C_{0} V}{e}=\frac{1}{2} +\frac{1}{2} \frac{eV-
U}{\sqrt{8t^{2}+(U-eV)^{2}}}, 
\label{charge.res}
\end{equation}
\begin{equation}
C_{\mu}-C_0=\frac{4e^{2}t^{2}}{[8t^{2}+(U-eV)^{2}]^{3/2}}.
\label{cap.res}
\end{equation}
Eqs.\ (7) and (8)
predict a charge transfer of $e$ across the 
chain and a capacitance peak at $eV=U$. 
Eq.\ (8) was obtained previously in Ref.\ \cite{markusandi},
where it was shown to describe charge transfer between two arbitrary
mesoscopic systems coupled weakly by tunneling.  Following the above
argument on the nature of charge transfer in the Mott insulator, one
may expect a result analogous to Eq.\ (8) to hold for larger
chains as well, since the effective coupling of the boundary dots via the 
intervening Mott-insulator should decrease exponentially with system size.
Indeed, the capacitance peaks at $eV\simeq U$ are found to become 
increasingly high and narrow (the area, which corresponds to the total
charge transferred, is conserved) as $L$ increases, but their shape is
found to be described very well by Eq.\ (8), with $t$ replaced
by an effective charge transfer matrix element $t_{\rm eff}$,
as shown in Fig.\
2(a).  Fitting the calculated capacitance to  Eq.\ (8), the
effective charge transfer matrix element is found to have the form
\begin{equation}
t_{\rm eff}\simeq t e^{-(L-2)/\xi(U/t)}
\label{teff}
\end{equation}
as shown in Fig.\ 2(b),
where the correlation length $\xi$ in the Mott-insulating phase of the 1D
Hubbard model is given by \cite{andyandi}
\begin{equation}
1/\xi(U/t) = 
 \frac{4t}{U} \int_{1}^{\infty} d y \, \frac{\ln (y + 
\sqrt{y^{2} - 1})}{\cosh (2 \pi t y/U)}.
\label{xinv}
\end{equation}
Eq.\ (9) indicates that the effective charge transfer matrix
element, which characterizes the resonant polarizability of the Mott
insulator, exhibits a finite-size scaling analogous to 
the conductivity of the system, which also decreases exponentially with
system size in the Mott insulator \cite{andyandi}.  $t_{\rm eff}$ is in fact
related to the equal-time Green's function, 
$t_{\rm eff}=tG(1,L)=
t\sum_{\sigma} \langle 0 | c^{\dagger}_{1\sigma} c_{L\sigma} 
| 0\rangle$, and it has already been argued \cite{andyandi} that
$G$ has the same scaling form as the conductivity for another choice
of boundary conditions.
Dielectric measurements thus 
present the intriguing possibility to study experimentally
the correlation length of a Mott insulator formed in a coherent
system of quantum dots.

While the charge transfer resonances in the Mott-insulating phase of the
model are well-described by the theory of Ref.\ \cite{markusandi}, it is
evident from Fig.\ 1 that the capacitance in the metallic phase
of the model, which may exhibit a low-bias double peak structure,
can not in general be described by an equation of the 
form of Eq.\ (8).  As shown in Fig.\ 3, this double peak structure
in the dilute metallic phase of the system is 
accentuated with increasing system size,  and corresponds to a
polarization charge with
well-defined plateaus quantized in units of $e/2$, unlike the integer charge
transfer
described by Eq.\ (7).  From Table I, we see
that the polarization of the system in the metallic phase proceeds via
the successive capture by the boundary dot of electrons from the Luttinger 
liquid states delocalized along the chain 
(since the antibound states are empty for $N<L$), the first at $eV=2t$ and
the second at $eV=eV_{2}\simeq 2U$.
The breakdown of Eqs.\ (7) and (8)
is due to the fact that the system can no longer be divided into just
two weakly coupled subsystems, as was assumed in Ref.\ \cite{markusandi}, 
and instead becomes one coherent whole in the metallic phase. 
 The fractional increments
of polarization charge shown in Fig.\ 3 arise because the trapping of
an electron from the gapless quantum states in the central part of
the array leads to an effective charge transfer over only half
the system.  
As a result, the magnitudes of the charges induced on the $1$-st and $L$-th 
dot are not equal to each other in the metallic phase in contrast 
to the equal and opposite charge polarization at both ends of 
a quantum dot chain in the insulating phase. For example, for $N=L-1$ 
electrons in the chain, the polarization charge on the $L$-th dot 
maximizes to one unit of charge $-e$ at $eV \approx 2t$, 
whereas the polarization charge on the 
$1$-st dot increases by $+e$ at $eV \approx 2U$.
The 
plateau structure in the capacitance near zero bias as seen in Fig.\ 3 
arises in the dilute quasiparticle/quasihole metallic limit 
($N \sim 1$ or $L-N \sim 1$), and its disappearance with increasing 
filling is an experimentally observable strong correlation effect in 
the charge polarization of the metallic phase. 

Let us comment on some of the idealizations employed in the above 
calculation.  The introduction of an interdot capacitance, 
neglected in Eq. (1), leads to
longer ranged site-off-diagonal interactions in the array, and a 
smoother distribution of
the externally applied voltage drop.  Such an extended 
Hubbard  model is no longer integrable 
via the Bethe ansatz technique,
but Lanczos direct diagonalization investigations \cite{kotlyarunp}
indicate that the physics is qualitatively similar to that described here.
Disorder, neglected in the present treatment, is not 
found to modify our main conclusions, as confirmed by our Lanczos
investigations \cite{kotlyarunp}.  The scaling form of the 
capacitance [Eqs.\ (8) and (9)] in the Mott insulating
phase of the system is preserved provided the disorder is not sufficiently
strong to lead to a compressible state, although the correlation length
$\xi$ is found to depend on disorder.  Similarly, the fractional polarization
charge plateaus shown in Fig.\ 3 are robust with respect to disorder, though
the voltage bias of the steps may be shifted.
We also remark that for temperatures $k_B T$ 
much larger than the effective charge transfer matrix element,
the form of the capacitance peaks given in Eq.\ (8) and
Fig.\ 3 will be replaced by a simple derivative of the Fermi function, of
width $k_B T$; however, the peak positions can still be used to distinguish
between metallic and Mott-insulating behavior. 

In conclusion, we have investigated coherent charge transfer in a 
strongly-correlated artificial linear molecule of 
tunnel-coupled quantum dots.  The polarizability
in the Mott-insulating phase of the system was found to exhibit a universal
scaling form analogous to the conductivity of the system, while the 
polarization charge in the dilute metallic phase of the system 
was found to be
quantized in units of $e/2$.
Equilibrium charge transfer measurements
present an intriguing alternative to transport measurements to characterize
the electronic states of ultrasmall structures. We believe that capacitance 
measurements carried out in carefully fabricated quantum dot chains should be 
able to observe the novel charge transfer resonances and the universal scaling 
behavior as well as the metallic charge fractionalization phenomenon  
predicted here.

This work is supported by the US-ONR.

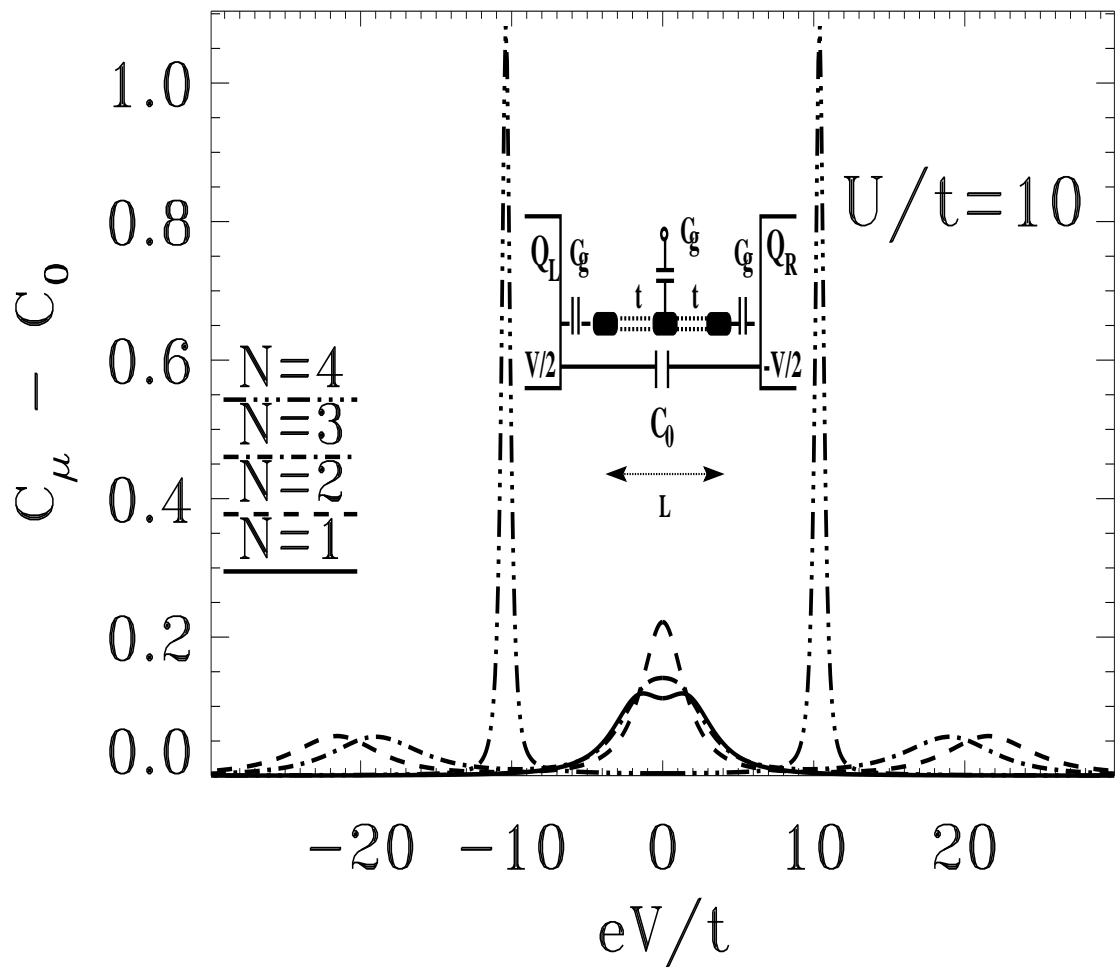
\begin{figure}
\caption{   
The quantum corrections to the capacitance plotted in units 
of $e^{2}/t$ of a chain of 4 quantum dots
as a function of the bias voltage $V$. The number of excess 
electrons in the chain and the interaction strength 
are as indicated in a legend.  
Inset: The equivalent circuit of the quantum dot array under study.}  
\end{figure}
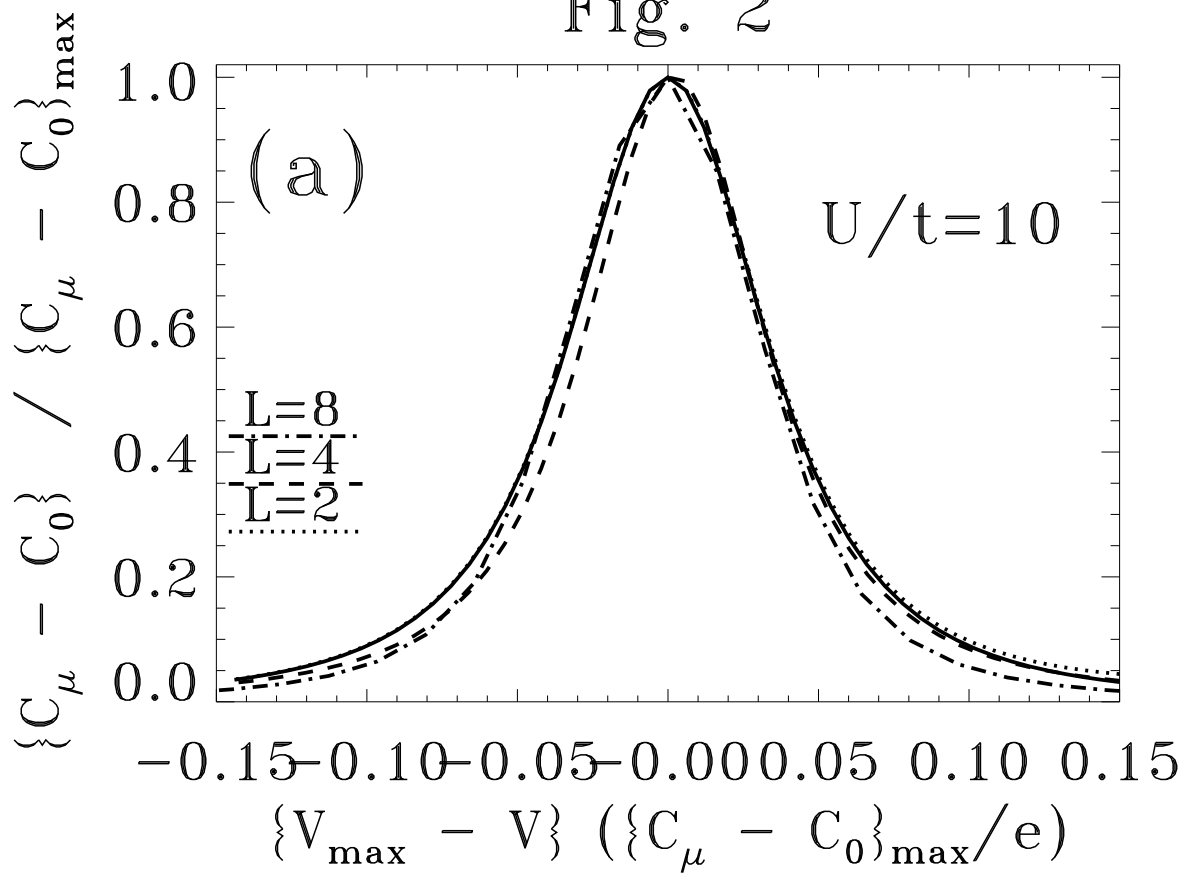
\begin{figure}
\caption{
(a) The charge transfer-induced resonant capacitance peaks for 
Mott-insulating chains of $2$, $4$, and 
$8$ dots (plotted as indicated in a legend), varying by several
orders of magnitude 
in height and width, are shown to collapse on the rescaled 
capacitance peak given by the Eq. (8) (solid line).
(b) The effective coupling $t_{eff}$ between the boundary dots of 
the Mott-insulating $L$-dot chains is plotted for $U/t=6,\ 7,\ 8,\ 9,\ 10$. 
The solid line with slope minus unity is shown to emphasize the scaling 
of the data given by the Eq. (9).}
\end{figure}
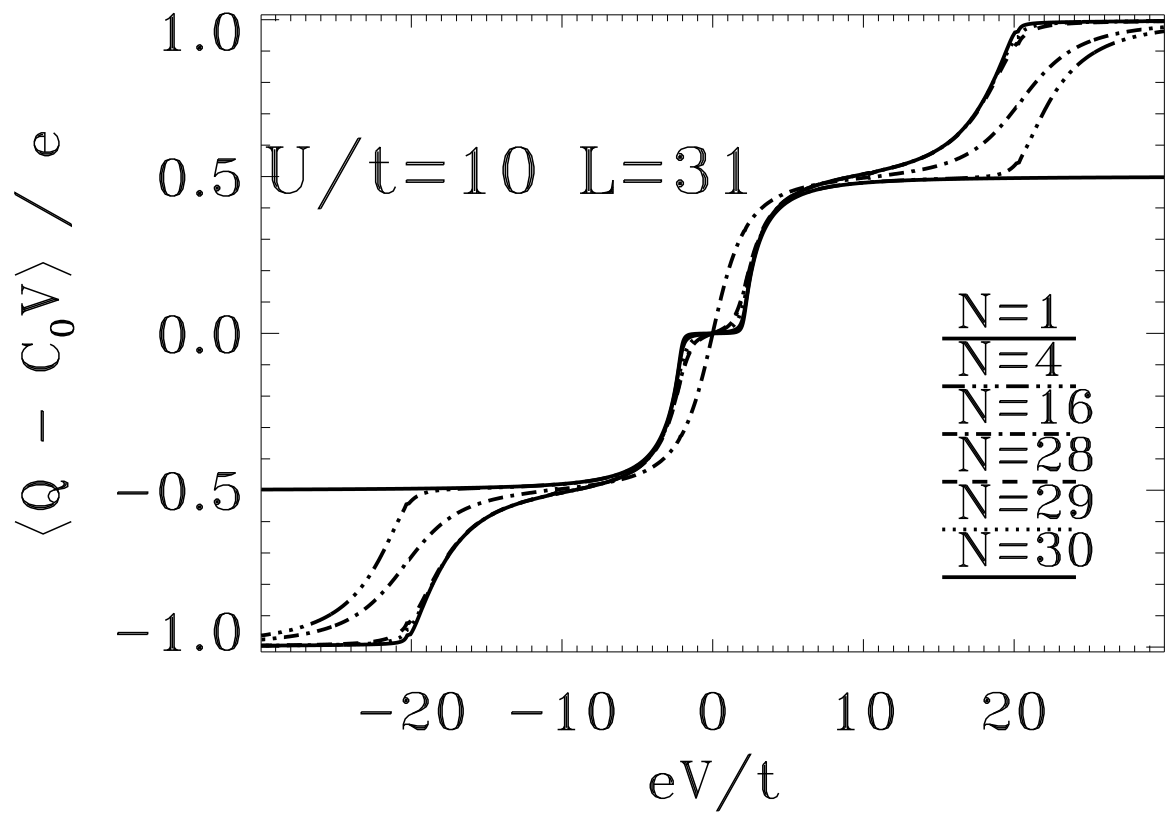
\begin{figure}
\caption{Quantum correction to the polarization charge induced on the 
external capacitor plates versus bias voltage for a
Hubbard chain of quantum dots in the metallic phase ($N < L$). Note that
$Q$ is quantized in units of $e/2$
($N \ll L$  or $N \stackrel{\displaystyle <}{\sim} L$)
reflecting a fractional ($e/2$) charge transfer
within the chain.} 
\end{figure}

\begin{table}
\begin{tabular}{|c|c|c|c|}
& I & II & III \\
& $2t/e < V < V_{1}$ & $V_{1} < V <  V_2$ & $V> V_{2}$ \\ \hline
& $k_{L}=i \ln(eV/2t)$ & $k_{L}=i \ln(eV/2t)$ & $k_{L}=i \ln(eV/2t)$ \\
$N=L$ & $k_{L-1}=\pi - i \ln(eV/2t)$ &
$k_{L-1}=\pi + i \sin^{-1} (i\sin k_{L}+U/2t)$ &
$k_{L-1}=- i \sin^{-1} (i\sin k_{L}+U/2t)$ \\
& & $\lambda_{M}=\sin k_{L}-i U/4t$ & $\lambda_{M}=\sin k_{L}-i U/4t$\\ \hline
& $k_{N}=i \ln(eV/2t)$ & $k_{N}=i \ln(eV/2t)$ & $k_{N}=i \ln(eV/2t)$ \\
$N<L$ & &  $\lambda_{M}=\sin k_{N}-i U/4t$ 
& $k_{N-1}=- i \sin^{-1} (i\sin k_{N}+U/2t)$ \\
& & & $\lambda_{M}=\sin k_{N}-i U/4t$\\ 
\end{tabular}

\caption{Complex roots of Eqs.\ (4) and (5)
(with exponential accuracy as $L\rightarrow \infty$)
corresponding to the bound and antibound states present in the ground state of
Eq.\ (1), as a function of the bias $V$.  Here
$eV_1= U/2+[(U/2)^2 + 4 t^2]^{1/2} $ and $eV_{2} = U+[U^2 + 4 t^2]^{1/2}$.
In the insulating phase of the model ($N=L$), polarization of the system
proceeds via transfer of the antibound state $k_{L-1}$ from one end of
the array to the other at the boundary of regions I and II.  In the metallic
phase, polarization of the system proceeds via the successive trapping
of electrons on the boundary dot with attractive potential at the onset
of regions I and III.}
\end{table}

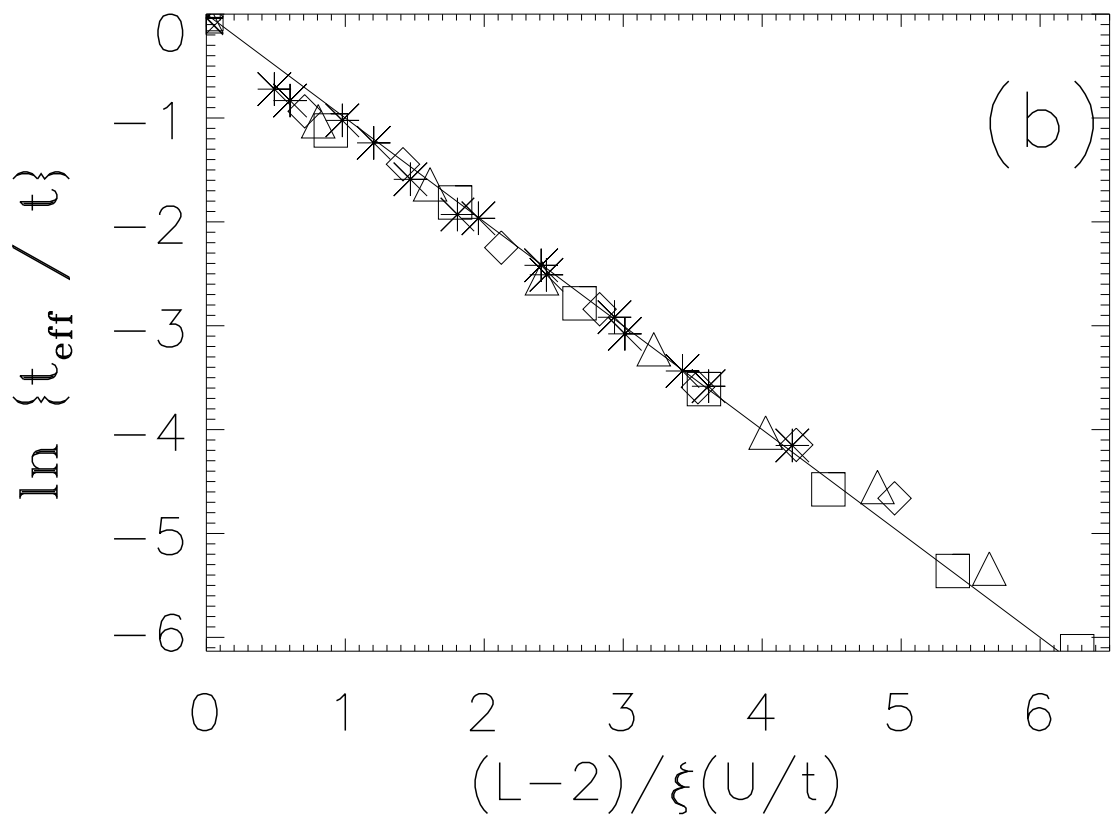
\end{document}